\documentclass[aps,prx,preprint,showpacs,superscriptaddress]{revtex4-1}
\usepackage{latexsym}
\usepackage{amssymb}
\usepackage{graphicx}
\usepackage{amsmath}
\usepackage{bm}
\usepackage[colorlinks,
          linkcolor=black,
            citecolor=black,
            urlcolor=blue
           ]{hyperref}
\usepackage{verbatim}
\usepackage{slashed}
\usepackage{mathrsfs}
\usepackage{extarrows}
\usepackage{mathtools}
\usepackage{comment}
\usepackage{mathtools,slashed}

\newcommand{\mH}{{\mathbb{H}}}
\newcommand{\mT}{{\mathbb{T}}}

\newcommand{\ver}{{\vec{r}}}
\newcommand{\cc}[2]{\left(\begin{array}{c}{#1}\\{#2}\end{array}\right)}

\begin{document}

\title{Dualization of ingappabilities through Hilbert-space extensions}

\author{Yuan Yao}
\email{smartyao@sjtu.edu.cn}
\affiliation{Institute of Condensed Matter Physics, School of Physics and Astronomy, Shanghai Jiao Tong University, Shanghai 200240, China}

\begin{abstract}
Typical dualities in arbitrary dimensions are understood through a Hilbert-space extension method.
By these results,
we rigorously dualize the quantum ingappabilities to discrete height model in one dimension which is inaccessible by earlier work such as flux-insertion arguments.
It turns out that the ingappabilities of quantum discrete height model is protected by an exotic ``modulating'' translation symmetry, 
which is a combination of modulating internal symmetry transformation and the conventional lattice translation.
It can be also generalize to higher-form gauge fields in arbitrary dimensions,
e.g., $\mathbb{Z}$-gauge theory in two dimensions with $\mathbb{Z}$ one-form symmetry and a modulating translation symmetry.

\end{abstract}

\maketitle
\tableofcontents

%%%%%%%%%%%%%%%%%%%%%%%%%%%%%%%%%%%%%%%%%%%%%%%%%%%%%%%%%%%%%%%%%%%%%%%%%%%%%%%

%%%%%%%%%%%%%%%%%%%%%%%%%%%%%%%%%%%%%%%%%%%%%%%%%%%%%%%%%%%%%%%%%%%%%%%%%%%%%%%%%%

\section{Introduction}
Understanding quantum many-body physics is an essential but complicated task in condensed matter and statistical physics,
due to complicated interactions and strong correlations.
Symmetry is a powerful tool to identify various quantum phases,
e.g., Landau-Wilson spontaneous symmetry breaking paradigm~\cite{Landau:1937aa} and symmetry-protected topological phases~\cite{Gu:2009aa,Chen:2010aa,Pollmann:2012aa,Wen:TOreview2013,Duivenvoorden:2013aa}.
%The concept of symmetry is also generalized in recent years to higher-form symmetry,
%higher-group symmetry, and there have been great progress in their affect on the low-energy physics and quantum phase classification.
Another useful approach to quantum many-body systems is duality,
which provides alternative viewpoint of ``the same'' physics,
where ``the same'' will be quantitatively clarified later by a unitary interpretation of duality.
Typical duality transformations are Kramers-Wannier (KW) duality, or its higher dimensional generalizations such as Abelian-Higgs (AH) duality and electric-magnetic (EM) duality.
Other exotic duality transformations are Kennedy-Tasaki transformation~\cite{Kennedy:1992aa,Kennedy:1992ab} and its generalizations~\cite{Oshikawa:1992aa,Li:2023aa,Choi:2024aa}.
In recent years,
duality transformations are understood as one type of the generalized symmetry, so-called non-invertible symmetry (See~\cite{Brennan:2023aa,Shao:2023aa,Carqueville:2023aa,Bhardwaj:2024aa,Schafer-Nameki:2024aa,Luo:2024aa} for reviews and references therein),
whose role in the quantum phase classifications or non-trivial constraints on low-energy physics has attracted a great number of efforts.

Quantum ingappability is another notable symmetry-associated non-perturbative concept in condensed matter physics;
one of the most important and typical example is the Lieb-Schultz-Mattis (LSM) theorem and its generalizations~\cite{Lieb:1961aa,Affleck:1986aa,OYA1997,Oshikawa:2000aa,Hastings:2004ab,NachtergaeleSims} which states that the system,
which respects U(1) and translation symmetry,  must be either gapped with degenerate ground states or gapless in the thermodynamic limit if the charge density is fractional.
However,
how this theorem is displayed if we view the system alternatively through a duality transformation is an open question.
One of the difficulty results from the noninvertibility of the duality transformation;
the spectrum of the dual theory can be completely different from the original theory before dualization.
The other related difficulty is the locality problem;
the lattice translation symmetry may not take a desirable form after dualization since duality transformation is generically not locality preserving.
In this work,
we dualize LSM-type theorem by a systematic extension of Hilbert space.
The Hilbert space is enlarged so that the duality transformation may become unitary~\cite{Li:2023aa}.
For KW, AH and EM dualities or their generalized analog,
the extended Hilbert space includes the symmetry twistings as a \textit{dynamical} degrees of freedom.
Thus,
the extended Hilbert space is a tensor product of the original Hilbert space and such artificial degrees of freedom.
The extension solves the first difficulty while the second difficulty of translation symmetry is overcome by a delicate extension of lattice translation symmetry in the extended Hilbert space so that it is reduced to a well-behaving,
e.g., unitary and locality preserving,
translation symmetry on the dual side after we go back to the physical Hilbert space of the dual theory eventually.
Thus a well-defined and -designed translation symmetry in the extended Hilbert space is a central intermediate result in this work,
although one might naively attempt to directly arrive at the dualized LSM theorem once the unitary transformation between two extended Hilbert spaces is obtained.

The main result of this work is various LSM-type ingappabilities of quantum generalized $\mathbb{Z}$-gauge theories in arbitrary dimensions.
These ingappabilities are protected by generalized internal $\mathbb{Z}$ symmetry~\cite{Gaiotto:2015aa,Gaiotto:2017aa} and an exotic ``modulating'' translation symmetry,
which is a combination of translation symmetry and an internal spatially modulating $\mathbb{Z}$-transformation.
The charge-filling condition in the original side is dualized into the modulation mode of this translation symmetry.

The paper is organized as follows.
In Sec.~\ref{kw},
we will carefully present the extended Hilbert space method in one dimension and discuss various possibilities of symmetry.
The LSM theorem in one dimension is dualized in the following Sec.~\ref{dual_LSM}.
Then,
higher dimensional statements will be studied subsequently,
and the general form of the duality in arbitrary dimensions are presented in Sec.~\ref{duality_arb}.
In Sec.~\ref{field_dual},
we also present how the traditional field-theoretical dualities can be reproduced in our framework.
In the Discussions~\ref{discussion},
we also discuss the noninvertibility of the symmetry in the physical Hilbert space.

\section{Duality transformations in one dimension}\label{kw}
In this section,
we first present Kramers-Wannier duality in one dimension distinguishing several cases.
We present how to enlarge the Hilbert space to make the duality to be a unitary transformation.
Since we will always assume the lattice is a spatial torus in its own dimension,
it is convenient,
e.g., in one dimension,
to use a modified $\delta$-function as
\begin{eqnarray}
\delta_{i,j}=\left\{\begin{array}{ll}1,&\text{ if }i=j\text{ mod } L;\\0,&\text{ otherwise.}\end{array}\right.
\end{eqnarray}

\subsection{$\mathbb{R}$-$\mathbb{R}$ duality: Warm-up}
This is the most flexible case where Hilbert space is
\begin{eqnarray}
|\{\phi_j\in\mathbb{R}\}_{j=1,\cdots,L}\rangle\otimes|\alpha\in\mathbb{R}\rangle
\end{eqnarray}
or its canonical momentum:
\begin{eqnarray}
|\{\pi_j\in\mathbb{R}\}_{j=1,\cdots,L}\rangle\otimes|p_\alpha\in\mathbb{R}\rangle
\end{eqnarray}
with the canonical relation $[\phi_j,\pi_k]=i\delta_{j,k}$ and $[\alpha,p_\alpha]=i$.
Of course,
the complete orthonormal basis could be also a mixture $|\{\pi_j\in\mathbb{R}\}_{j=1,\cdots,L}\rangle\otimes|\alpha\in\mathbb{R}\rangle$.
In this case,
we do not have U$(1)$ symmetry but $\mathbb{R}$ symmetry generated by
\begin{eqnarray}
\prod_j\exp(i\theta\pi_j),\,\,\theta\in\mathbb{R}.
\end{eqnarray}
The translation symmetry is extended to
\begin{eqnarray}
&&T\phi_{j}T^{-1}=\left\{\begin{array}{ll}\phi_{j+1},&\text{ if }j\neq L,\\
\phi_1-\alpha,&\text{ if }j= L,\end{array}\right.\\
&&T\pi_j T^{-1}=\pi_{j+1},\\
&&T\alpha T^{-1}=\alpha,
\end{eqnarray}
Therefore,
$\alpha$ \textit{effectively} twists the boundary condition if we treat it as a background field.
The translation operator $T$ can be fully determined after its effect on $p_\alpha$ is defined,
as we will consider later.
For convenience,
we define a link variable field:
\begin{eqnarray}
[\alpha]_{j+1/2}=\alpha\delta_{j,L}.
\end{eqnarray}

The dual part is proposed as a height model on the dual lattice chain
\begin{eqnarray}
|\{h_{j-1/2}\in\mathbb{R}\}_{j=1,\cdots,L}\rangle\otimes|\beta\in\mathbb{R}\rangle
\end{eqnarray}
or its canonical momentum:
\begin{eqnarray}
|\{\pi_{h,j-1/2}\in\mathbb{R}\}_{j=1,\cdots,L}\rangle\otimes|p_\beta\in\mathbb{R}\rangle
\end{eqnarray}
with the canonical relation $[h_{j-1/2},\pi_{h,k-1/2}]=i\delta_{j,k}$ and $[\beta,p_\beta]=i$.
We also define a link variable on the dual lattice:
\begin{eqnarray}
[\beta]_{j}=\beta\delta_{j,L}.
\end{eqnarray}
We note that the link of the original lattice is labelled by the site of the dual lattice.
Thus,
in one dimension,
we simply identify these two coordinate systems.
It enables us to define a discrete version of exterior derivative which maps a site variable of one lattice to a link variable of its dual:
\begin{eqnarray}
[\Delta\phi]_{j+1/2}=\phi_{j+1}-\phi_j,
\end{eqnarray}
or vice versa:
\begin{eqnarray}
[\Delta h]_j=h_{j+1/2}-h_{j-1/2}.
\end{eqnarray}

We motivate the duality by an operator mapping:
\begin{eqnarray}
&&KW[\Delta\phi-\alpha]KW^\dagger=\pi_{h},\\
&&KW [\pi]KW^\dagger=\Delta h-\beta,\\
&&KW (p_\alpha)KW^\dagger=\phi_L.
\end{eqnarray}
In the following discussions,
we will not distinguish the link field $\alpha$ and the operator $\alpha$ since it should be clear by the context.

The above three relations fully characterize $KW$ since we can solve them out as
\begin{eqnarray}\label{1_d_phi}
\left\{\begin{array}{l}KW\alpha KW^\dagger=-\sum_k\pi_{h,k+1/2};\\
KW p_\alpha KW^\dagger=h_{1/2};\\
KW \phi_j KW^\dagger=p_\beta+\sum_{k=1}^{j}\pi_{h,k+1/2}-\delta_{j,L}\sum_{k}\pi_{h,k+1/2},\,\,(j=1,\cdots,L);\\
KW \pi_jKW^\dagger=h_{j+1/2}-h_{j-1/2}-\beta\delta_{j,L},\end{array}\right.
\end{eqnarray}
and its inverse
\begin{eqnarray}\label{1_d_h}
\left\{\begin{array}{l}KW^\dagger\beta KW=-\sum_k\pi_{k};\\
KW^\dagger p_\beta KW=\phi_L;\\
KW^\dagger h_{j+1/2} KW=p_\alpha+\sum_{k=1}^{j}\pi_{k}-\delta_{j,L}\sum_{k}\pi_{k},\,\,(j=1,\cdots,L);\\
KW^\dagger\pi_{h,j+1/2}KW=\phi_{j+1}-\phi_j-\alpha\delta_{j,L},\end{array}\right.
\end{eqnarray}
Alternatively,
its action on the Hilbert space is:
\begin{eqnarray}
KW|\{\pi_j\}\rangle\otimes|p_\alpha\rangle=|\{h_{j+1/2}\}\rangle\otimes|\beta\rangle,
\end{eqnarray}
where $h_{j+1/2}$'s and $\beta$ are given in the above inverse transformation.

One should note it that the current duality transformation is an \textit{isomorphism} and invertible transformation as long as we extend the original Hilbert space to include the ``twisting'' or its canonical conjugate as a dynamical degrees of freedom.

\subsection{U(1)-$\mathbb{Z}$ duality: Rigidity but richness}
We consider a more rigid situation than the $\mathbb{R}$-$\mathbb{R}$ duality in the sense of the Hilbert space;
now $\phi_j$'s are angle-valued in the
so-called quantum rotor model (or ``quantum classical XY model'') tensored with its $\alpha$-extension:
\begin{eqnarray}
|\{\exp(i\phi_j)\in\text{U}(1)\}_{j=1,\cdots,L}\rangle\otimes|\exp(i\alpha)\in \text{U}(1)\rangle,
\end{eqnarray}
or its canonical conjugate:
\begin{eqnarray}
|\{\pi_j\in\mathbb{Z}\}_{j=1,\cdots,L}\rangle\otimes|p_\alpha\in\mathbb{Z}\rangle.
\end{eqnarray}
The dual side is a discrete height model:
\begin{eqnarray}
|\{h_{j-1/2}\in\mathbb{Z}\}_{j=1,\cdots,L}\rangle\otimes|\beta\in\mathbb{Z}\rangle
\end{eqnarray}
or by its canonical momentum which is U$(1)$-valued:
\begin{eqnarray}
|\{\exp(i\pi_{h,j})\in\text{U}(1)\}_{j=1,\cdots,L}\rangle\otimes|\exp(ip_\beta)\in\text{U}(1)\rangle.
\end{eqnarray}

Thence,
the duality turns out to be exponentiated properly from the earlier transformation:
\begin{eqnarray}
\left\{\begin{array}{l}KW\exp(i\alpha) KW^\dagger=\exp(-i\sum_k\pi_{h,k});\\
KW p_\alpha KW^\dagger=h_{1/2};\\
KW \exp(i\phi_j) KW^\dagger=\exp(ip_\beta)\exp(i\sum_{k=1}^j\pi_{h,k+1/2}-i\delta_{j,L}\sum_k\pi_{h,k+1/2}),\,\,(j=1,\cdots,L);\\
KW \pi_jKW^\dagger=h_{j+1/2}-h_{j-1/2}-\beta\delta_{j,L},\end{array}\right.
\end{eqnarray}
and its inverse
\begin{eqnarray}
\left\{\begin{array}{l}KW^\dagger\beta KW=-\sum_k\pi_{k};\\
KW^\dagger \exp(ip_\beta) KW=\exp(i\phi_L);\\
KW^\dagger h_{j+1/2} KW=p_\alpha+\sum_{k=1}^j\pi_{k}-\delta_{j,L}\sum_k\pi_k,\,\,(j=1,\cdots,L);\\
KW^\dagger\exp(i\pi_{h,j+1/2})KW=\exp[i(\phi_{j+1}-\phi_j-\alpha\delta_{j,L})].\end{array}\right.
\end{eqnarray}
It is obvious how to do various exponentiations once we know the most flexible transformation and its inverse in Eqs.~(\ref{1_d_phi},\ref{1_d_h}); once we meet an angle-valued variable,
we take its exponentiation to make it well-defined.

\subsection{$\mathbb{Z}_n$-$\mathbb{Z}_n$ duality}
When we restrict the angle $\phi$ to a $\mathbb{Z}_n$-clock,
the U(1)-U(1) duality above becomes the conventional Kramers-Wannier duality.
To be conventional,
the degrees of freedom will be rewritten $(2\pi\equiv1)$:
\begin{eqnarray}
\left\{\begin{array}{l}\exp(i\phi_j)\mapsto\sigma_j;\\\exp(i\pi_j)\mapsto\tau_j;\\
\exp(i\alpha)\mapsto a;\\
\exp(ip_\alpha)\mapsto p_a,
\end{array}\right.
\end{eqnarray}
with
\begin{eqnarray}
\sigma_j^n=\tau_j^n=a^n=p_a^n=1;\,\tau_k\sigma_j=\sigma_j\tau_k\exp\left(\frac{i2\pi}{n}\delta_{j,k}\right);\,p_aa=ap_a\exp\left(\frac{i2\pi}{n}\right).
\end{eqnarray}

Similarly for the dual side:
\begin{eqnarray}
\left\{\begin{array}{l}\exp(ih_{j-1/2})\mapsto\mu_j;\\\exp(i\pi_{h,j-1/2})\mapsto\lambda_j;\\
\exp(i\beta)\mapsto b;\\
\exp(ip_\beta)\mapsto p_b,
\end{array}\right.
\end{eqnarray}
with
\begin{eqnarray}
\mu_j^n=\lambda_j^n=b^n=p_b^n=1;\,\lambda_k\mu_j=\mu_j\lambda_k\exp\left(\frac{i2\pi}{n}\delta_{j,k}\right);\,p_bb=bp_b\exp\left(\frac{i2\pi}{n}\right).
\end{eqnarray}
Thence,
the duality turns out to be:

\begin{eqnarray}
\left\{\begin{array}{l}KWa KW^\dagger=\prod_k\lambda_k^{-1};\\
KW p_a KW^\dagger=\mu_L;\\
KW \sigma_j KW^\dagger=p_b\prod_{k=1}^j\lambda_k\left(\prod_{k}\lambda_k\right)^{-\delta_{j,L}},\,\,(j=1,\cdots,L);\\
KW \tau_jKW^\dagger=\mu_{j+1}b^{-\delta_{j,L}}\mu^{-1}_j,\end{array}\right.
\end{eqnarray}
and its inverse
\begin{eqnarray}
\left\{\begin{array}{l}KW^\dagger b KW=\prod_k\tau^{-1}_{k};\\
KW^\dagger p_b KW=\sigma_L;\\
KW^\dagger \mu_j KW=a\prod_{k=1}^j\pi_k\left(\prod_k\pi_k\right)^{-\delta_{j,L}},\,\,(j=1,\cdots,L);\\
KW^\dagger\lambda_jKW=\sigma_{j+1}a^{-\delta_{j,L}}\sigma_j^{-1}.\end{array}\right.
\end{eqnarray}

\section{Dualized Lieb-Schultz-Mattis theorem}\label{dual_LSM}
In this section,
we will first prove the LSM theorem for quantum rotor model and dualize it as an ingappability for quantum discrete height model.

\subsection{Lieb-Schultz-Mattis theorem for quantum rotor model}

First,
we will give a twisting-operator approach to Lieb-Schultz-Mattis (LSM) theorem for quantum rotor model in one dimension.
The local Hilbert space at each lattice site is an angle variable $\phi_j\sim\phi_j+2\pi$, which enables us to use
\begin{eqnarray}
|\{\exp(i\phi_j)\in\text{U}(1)\}_{j=1,\cdots,L}\rangle,
\end{eqnarray}
as a faithful and complete description of the entire Hilbert space,
in addition to its canonical and equivalently complete correspondence:
\begin{eqnarray}
|\{\pi_j\in\mathbb{Z}\}_{j=1,\cdots,L}\rangle,
\end{eqnarray}
where the ``angular momentum'' satisfies
\begin{eqnarray}
[\pi_j,\exp(i\phi_k)]=\delta_{j,k}.
\end{eqnarray}
Unlike spin models or fermionic systems,
the spectrum of local operator $\pi_j$ is not bounded,
so it is useful to introduce a many-body concept of thermodynamic limit:

{\bf Thermodynamic limit: }An eigenstate $|\Psi_L\rangle$ of a Hamiltonian in one dimension of length $L$ has a thermodynamic limit if any operator polynomial
\begin{eqnarray}
\mathcal{F}[\pi_{j-l},\exp(\pm i\phi_{j-l}),\cdots,\pi_{j+l},\exp(\pm i\phi_{j+l})]
\end{eqnarray}
has a well-defined expectation value $\lim_{L\rightarrow\infty}\langle\Psi_L|\mathcal{F}|\Psi_L\rangle$ as $L\rightarrow\infty$,
where $l$ is the maximal interaction range of the Hamiltonian.

We expect this property should be satisfied by generic physical systems with a bounded interaction range since it reflects the extensibility.

We state the theorem as follows:

{\bf Theorem 1 --- }\textit{LSM theorem of quantum rotor models in one dimension:}
If a quantum rotor chain respects lattice translation symmetry $T$:
\begin{eqnarray}
T\exp(i\phi_j)T^{-1}=\exp(i\phi_{j+1});\,\,T\pi_jT^{-1}=\pi_{j+1},
\end{eqnarray}
under periodic boundary condition (PBC),
and U$(1)$-rotational symmetry generated by
\begin{eqnarray}
\prod_{j=1}^L\exp(i\theta\pi_j),\,\,\theta\in[0,2\pi),
\end{eqnarray}
then,
as $L\rightarrow\infty$, there must exist multiple lowest-lying energy eigenstates \textit{within} a fixed U$(1)$-charge sector with $p/q$-fractional charge per unit cell
\begin{eqnarray}
Q=\sum_{j=1}^L\pi_j=\frac{p}{q}L,
\end{eqnarray}
as long as one of the lowest-lying states has a well-defined thermodynamic limit of U$(1)$-symmetric polynomials of local operators (while we do not require the full thermodynamic limit).

\textit{Proof: }

We prove it by contradiction;
we assume that the lowest-lying $E_0$-energy eigenstate $|\text{G.S.}\rangle$ in the $(p/q)$ U(1)-charge density sector is unique and has a finite energy gap below the excited states within the same charge sector.
Thus $|\text{G.S.}\rangle=T|\text{G.S.}\rangle$ up to a phase.
We make a trial state:
\begin{eqnarray}
|\Phi\rangle\equiv\exp\left(\sum_{j=1}^L\frac{i2\pi j}{L}\pi_j\right)|\text{G.S.}\rangle,
\end{eqnarray}
which is also in the same charge sector because the twisting operator commutes with U$(1)$.
Let us examine the energy difference:
\begin{eqnarray}
&&\langle\Phi|H|\Phi\rangle-E_0\nonumber\\
&=&\frac{2\pi}{L}\left[\sum_{j=1}^{L}\left(\langle\text{G.S.}|j[\pi_j,H]|\text{G.S.}\rangle\right)-\frac{L+1}{2}\left(\langle\text{G.S.}|[\sum_{k=1}^L\pi_k,H]|\text{G.S.}\rangle\right)\right]+O(1/L)\nonumber\\
&=&\frac{2\pi}{L}\left[\sum_{j=1}^{L}\left(\langle\text{G.S.}|j[\pi_1,H]|\text{G.S.}\rangle\right)-\frac{L+1}{2}\left(\langle\text{G.S.}|[\sum_{k=1}^L\pi_1,H]|\text{G.S.}\rangle\right)\right]+O(1/L)\nonumber\\
&=&0,\text{ as }L\rightarrow\infty,
\end{eqnarray}
where we have used the fact that the Hamiltonian commutes with the U(1) generator $\sum_k\pi_k$, and $|\text{G.S.}\rangle$ has a thermodynamic limit to obtain $O(1/L)$ estimate in the first line since the operators contributing to $O(1/L)$ is U$(1)$-symmetric.
In the second line,
we use the translation symmetry.
Furthermore,
$|\Phi\rangle$ has a different lattice momentum as $|\text{G.S.}\rangle$'s $\exp(iP_0)$ as a contradiction:
\begin{eqnarray}
T|\Phi\rangle&=&\left[T\exp\left(\sum_{j=1}^L\frac{i2\pi j}{L}\pi_j\right)T^{-1}\right]T|\text{G.S.}\rangle\nonumber\\
&=&\exp\left(iP_0-i2\pi \frac{p}{q}\right)|\Phi\rangle,
\end{eqnarray}
where we have used the fact that $\exp(i2\pi\pi_1)=1$ as an operator equation.

\subsection{Dualized LSM theorem}
To simplify the notations,
we will move the dual lattice along -$x$ axis by 1/2,
i.e., $h_{j+1/2}\mapsto h_j$ \textit{etc.},
to completely overlap with the original lattice.

We first state the dual LSM theorem as follows:

{\bf Theorem $\check{1}$ --- }\textit{LSM theorem of quantum $\mathbb{Z}$-height chains --- }
If a quantum $\mathbb{Z}$-height Hamiltonian respects ``modulating'' lattice translation symmetry $T_{p/q}$:
\begin{eqnarray}\label{mod_transl_1}
T_{p/q}h_jT_{p/q}^{-1}=h_{j+1}+p\delta_{j=1\text{ mod }q};\,\,T_{p/q}\pi_{h,j}T_{p/q}^{-1}=\pi_{h,j+1},
\end{eqnarray}
under periodic boundary condition (PBC),
and an onsite $\mathbb{Z}$-raising symmetry generated by
\begin{eqnarray}
\prod_{j=1}^L\exp(im\pi_{h,j}),\,\,m\in\mathbb{Z},
\end{eqnarray}
then,
as $L\rightarrow\infty$, there must exist multiple lowest-lying energy eigenstates \textit{within any} $\mathbb{Z}$-symmetry charge Hilbert subspace
as long as one of the lowest-lying states has a thermodynamic limit of $\mathbb{Z}$-symmetric polynomials of local terms.

{\bf Remark:} The condition ``any $\mathbb{Z}$-symmetry charge Hilbert subspace'' actually strengthens the theorem than without such a restriction to this Hilbert subspace.

\textit{Proof: }

Let us denote the Hamiltonian of the quantum height model as 
\begin{eqnarray}
\check{H}[\{h_{j+1}-h_j-p\delta_{j=1\text{ mod }q},\pi_{h,j}\}]
\end{eqnarray}
whose $\{h_{j+1}-h_j\}$-dependence is due to its $\mathbb{Z}$-raising symmetry while the additional (trivial) ``$-p\delta_{j=1\text{ mod }q}$'' is for later notational convenience.
Then we do a unitary transformation so that $U_{p/q}\pi_{h,j}U^\dagger_{p/q}=\pi_{h,j}$ and
\begin{eqnarray}
U_{p/q}(h_{j+1}-h_{j}-p\delta_{j=1\text{ mod }q}) U_{p/q}^\dagger=\left\{\begin{array}{ll}h_{j+1}-h_j,&\text{ if }j\neq L,\\
h_{1}-h_{L}-\frac{p}{q}L,&\text{ if }j=L.\end{array}\right.
\end{eqnarray}
to obtain a ``gauge''-equivalent Hamiltonian:
\begin{eqnarray}
\check{H}_{pL/q}&\equiv&U_{p/q}\check{H}U^\dagger_{p/q}\nonumber\\
&=&\check{H}[\{h_{j+1}-h_j-\frac{pL}{q}\delta_{j,L},\pi_{h,j}\}].
\end{eqnarray}
Thus,
such a unitary transformation effectively accumulates all the twisting at the bond between the sites $L$ and $1$.
Indeed,
the modulating translation symmetry becomes
\begin{eqnarray}\label{transl_1d}
&&T\equiv T_{p/q}U^\dagger_{p/q},\\
&&[\check{H}_{pL/q},T]=0,
\end{eqnarray}
and
\begin{eqnarray}
Th_{j}T^{-1}=h_{j+1}-\frac{pL}{q}\delta_{j,L}.
\end{eqnarray}

Then we extend the original physical Hilbert space $|\{h_j\}_{j=1,\cdots,L}\rangle$ to
\begin{eqnarray}
|\{h_j\}_{j=1,\cdots,L}\in\mathbb{Z}\rangle\otimes|\beta\in\mathbb{Z}\rangle,
\end{eqnarray}
where the canonical-to-$\beta$ momentum $\exp(ip_\beta)\in$U(1) is circle-valued.
We define a Hamiltonian $\check{\mathbb{H}}$ in this artificial larger Hilbert space such that
\begin{eqnarray}
\check{\mathbb{H}}|(\cdots)\rangle\otimes|\beta\rangle=\check{H}_\beta|(\cdots)\rangle\otimes|\beta\rangle,
\end{eqnarray}
in which one should note it that there should be an explicit appearance of the new operator $\hat{\beta}$ in $\mH$ while $\beta$ without ``hat'' is an integer.
In the following,
we will sometimes abuse these two notations,
which should be clear by the context.

Clearly,
$[\check{\mathbb{H}},\hat{\beta}]=[\check{\mathbb{H}},\prod_{j}\exp(-i\pi_{h,j})]=0$,
the Hilbert space can be split into energy eigenstates with their own $\hat{\beta}$- and $\prod_{j}\exp(-i\pi_{h,j})$-eigenvalues:
\begin{eqnarray}
\check{\mathbb{H}}|\Psi:\beta,\exp(i\alpha)\rangle=E|\Psi:\beta,\exp(i\alpha)\rangle
\end{eqnarray}
so the original ingappability to be proven is the ingappability problem in the $\hat{\beta}=pL/q$-subspace of the current extended Hilbert space.
%Thus,
%we will suppress ``hat''s to imply that we have restricted ourselves in that certain subspace.

Then we can prove by contradiction as follows.
We assume that $\check{\mathbb{H}}$ has a unique gapped lowest-lying state within the $\hat{\beta}=pL/q,\,\prod_{k}\exp(-i\pi_{h,k})=\exp(i\alpha)$-Hilbert subspace.
Then we do a Kramers-Wannier transformation $KW^\dagger$ and the ingappability problem is equivalently transformed to that of $KW^\dagger \check{\mathbb{H}}KW$ within the Hilbert subspace with $\exp(i\hat{\alpha})=\exp(i\alpha)$ and U$(1)$-charge density $\sum_{j}\pi_j/L=-p/q$.
Therefore,
once we could prove the translation symmetry on the \textit{extended} rotor-model side is reduced to (or unitarily equivalent to) the conventional lattice translation, and the dual side also has well-defined thermodynamic limit,
we complete the proof by referring to {\bf Theorem 1}.

First,
let us investigate the form the translation on the rotor side.
To do so,
we find that the operator $T$ in Eq.~(\ref{transl_1d}) in the original physical Hilbert space can be realized in the extended space as
\begin{eqnarray}
\left\{\begin{array}{l}\mT\hat{\beta} \mT^{-1}=\hat{\beta};\\
\mT\exp(ip_\beta)\mT^{-1}=\exp(ip_\beta)\exp(i\pi_{h,L});\\
\mT h_{j}\mT^{-1}=h_{j+1}-\hat{\beta}\delta_{j,L};\\
\mT\exp(i\pi_{h,j})\mT^{-1}=\exp(i\pi_{h,j+1}),\end{array}\right.
\end{eqnarray}
which exists since it is consistent with all the canonical relations.
One might wonder whether other translation $\mT$ rule of $\exp(ip_\beta)$ that could reduce to $T$ in the (pre-extended) physical Hilbert space can be used.
Although $\mT\exp(ip_\beta)\mT^{-1}$ is ``invisible'' in the physical Hilbert space,
its form will essentially change the translation symmetry on the dual side;
our goal is to make $\mT$ reduce to lattice translation on the dual side as well,
which needs a delicate designation as above.
Indeed,
on the dual side,
this translation symmetry becomes $\mT\circ KW$ as
\begin{eqnarray}
\left\{\begin{array}{l}(\mT\circ KW)\exp(i\hat{\alpha})(\mT\circ KW)^{-1}=\exp(i\hat{\alpha});\\
(\mT\circ KW)p_\alpha(\mT\circ KW)^{-1}=p_\alpha+\pi_1;\\
(\mT\circ KW)\exp(i\phi_j)(\mT\circ KW)^{-1}=\exp(i\phi_{j+1})\exp(i\hat{\alpha}\delta_{j,L});\\
(\mT\circ KW)\pi_j(\mT\circ KW)^{-1}=\pi_{j+1}.
\end{array}\right.
\end{eqnarray}
%We do not need to care about the transformation of $p_\alpha$ because 
%$KW^\dagger \check{\mH} KW$ does not depend on the operator $p_\alpha$,
%but we only need to know its existence.
Then,
when we go to $\exp(i\hat{\alpha})=\exp(i\alpha)$ sector,
we find that the translation $\mT\circ KW$ is indeed unitarily equivalent to the conventional translation symmetry up to a unitary transformation that commutes with and preserves $\exp(i\hat{\alpha})$ and $\sum_j\pi_j$.

Now,
we proceed to the proof of the thermodynamic limit of U$(1)$-symmetric operators on the rotor side.
First,
we note that any U$(1)$-symmetric operator can be written into the functional form of
\begin{eqnarray}
\mathcal{F}\{\exp[i(\phi_{j+1}-\phi_j-\alpha\delta_{j,L})],\pi_j\},
\end{eqnarray}
then the well-defined thermodynamic limit of it is equivalent to that of
\begin{eqnarray}
KW\circ\mathcal{F}\{\exp[i(\phi_{j+1}-\phi_j-\hat{\alpha}\delta_{j,L}),\pi_j\}\circ KW^\dagger=\mathcal{F}\{\exp[i\pi_{h,j},h_{j+1}-h_j-\hat{\beta}\delta_{j,L}\},
\end{eqnarray}
which is $\mathbb{Z}$-symmetric thereby having well-defined thermodynamic limit by the condition of the {\bf Theorem $\check{1}$}.
Thus,
the proof of {\bf Theorem $\check{1}$} is completed.

\section{Duality transformations in two dimensions}
We show how the Kramers-Wannier transformation is generalized towards higher dimensions. 
For simplicity,
we consider two-dimensional spatial lattice first.

We label the square lattice site by $\vec{r}$ where scalar-type operators stay.
Vector fields stay on the oriented links,
e.g., the link starting from $\vec{r}_1$ to a neighboring $\vec{r}_2$ labelled by $(\vec{r}_1,\vec{r}_2)$.
In addition,
we can also define a dual of the lattice,
e.g., by ``$\vec{r}+x/2+y/2$''  where $x$ and $y$ denote unit vectors.
Conventionally,
$A(\ver_1,\ver_2)=-A(\ver_2,\ver_1)$.
Sometimes,
we will directly write down ``$A(t)$'' with ``$t$'' an oriented link.
Furthermore,
we can have field defined on plaquettes which are labelled by its center coordinate.
In analog to the exterior derivatives,
we define $\Delta$ which can map a link vector field to a plaquette field,
for instance,
\begin{eqnarray}
&&[\Delta A](\vec{r}+x/2+y/2)\nonumber\\
&=&A(\ver,\ver+x)+A(\vec{r}+x,\vec{r}+x+y)+A(\vec{r}+x+y,\vec{r}+y)+A(\ver+y,\ver),
\end{eqnarray}
and map a scalar field to a vector field:
\begin{eqnarray}
[\Delta \phi](\ver,\ver+\mu)=\phi(\ver+\mu)-\phi(\ver),\,\,(\mu=x,y).
\end{eqnarray}

\begin{figure}[t]
\centering
\includegraphics[width=8.8cm,pagebox=cropbox,clip]{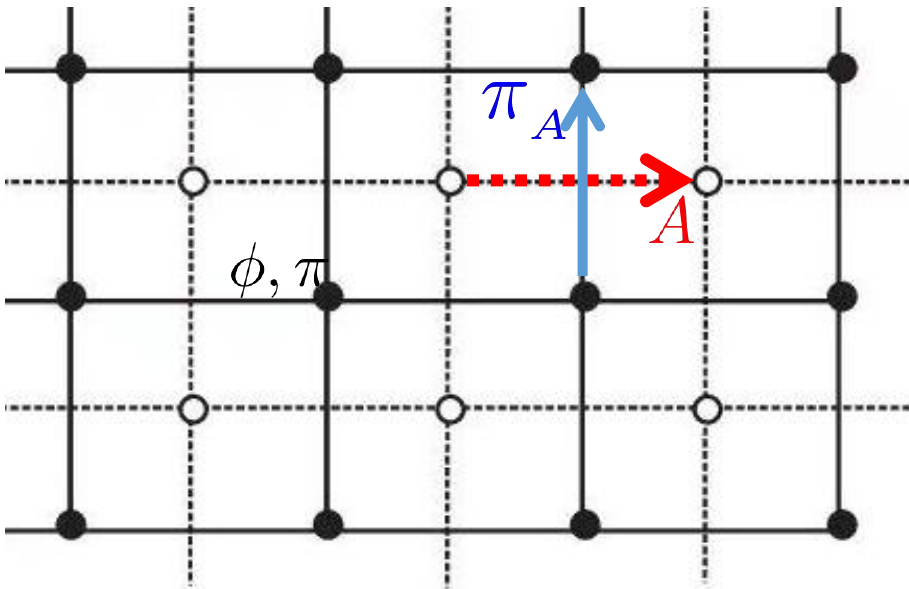}
\caption{Black (White) dots form the (dual) lattice where various degrees of freedom are defined.}
\label{DOF}
\end{figure}

These degrees of freedom are shown in FIG.~\ref{DOF}.
We will give several types of dualities and starting from the most flexible one.

\subsection{$\mathbb{R}$-$\mathbb{R}$ duality}
We first describe the Hilbert spaces separately and then formulate the duality as a unitary transformation between them.
\subsubsection{$\mathbb{R}$-scalar}
This model is a natural generalization of the one-dimensional case,
and the extended Hilbert space is
\begin{eqnarray}
|\{\phi(\vec{r})\in\mathbb{R}\}\rangle\otimes|(\alpha_x,\alpha_y)\in\mathbb{R}\times\mathbb{R}\rangle,
\end{eqnarray}
with their canonical conjugate:
\begin{eqnarray}
|\{\pi(\vec{r})\in\mathbb{R}\}\rangle\otimes|(p_\alpha^x,p_\alpha^y)\in\mathbb{R}\times\mathbb{R}\rangle,
\end{eqnarray}
where
\begin{eqnarray}
[\alpha_\mu,p_\alpha^\nu]=i\delta^\nu_\mu.
\end{eqnarray}

\subsubsection{$\mathbb{R}$-vector}
We define the vector fields on the dual links for the sake of notational convenience of duality later:
\begin{eqnarray}
\frac{|\{A(\ver+x/2+y/2,\ver+x/2+y/2+\mu)\in\mathbb{R}\}\rangle}{A\sim A+\Delta f:f(\ver)\in\mathbb{R}}\otimes|N\in\mathbb{R}\rangle,
\end{eqnarray}
where $N$ will play the role of twisting of boundary condition of vector field associated with 1-form $\mathbb{R}$-symmetry.
The gauge equivalence is inconvenient,
so we use a gauge-invariant but faithful representation by Wilson loop operators:
\begin{eqnarray}
\begin{array}{l}|\{\sum_{\vec{t}\in\check{l}}A(\vec{t})\in\mathbb{R}\}_{\check{l}}\rangle\otimes|N\in\mathbb{R}\rangle,\end{array}
\end{eqnarray}
%where $\check{l}_\mu$ is a closed loop winding only around $\mu$ axis once,
where $\check{l}$ is a closed loop
in the dual lattice.
Another useful representation is
\begin{eqnarray}
\begin{array}{l}|\{\sum_{\vec{t}\in\check{L}_\mu}A(\vec{t})\in\mathbb{R}\}_{\mu=x,y},\{\Delta A\in\mathbb{R}\}\rangle\otimes|N\in\mathbb{R}\rangle,\end{array}
\end{eqnarray}
where $\check{L}_\mu$ is a special closed loop winding only around $\mu$ axis once,
of the dual lattice sites and it transversely intersects the boundary links as shown in FIG.~\ref{phi_dual}.
The canonical momentum of $A$'s is, e.g., electric fields in the Maxwell theory.
However,
we will use another convention and define their momenta on the link of the original lattice:
\begin{eqnarray}
\left\{\begin{array}{l}\left[A(\ver+x/2+y/2,\ver+x/2+y/2+x),\pi_A(\ver+x,\ver+x+y)\right]=i;\\
\left[A(\ver+x/2+y/2,\ver+x/2+y/2+y),\pi_A(\ver+x+y,\ver+y)\right]=i,\end{array}\right.
\end{eqnarray}
or more compactly,
\begin{eqnarray}
\left[A(\check{l}),\pi_A(t)\right]=i\text{Int}(\check{l},t),
\end{eqnarray}
where Int$(\check{l},t)$ of two oriented links is nonzero only if $\check{l}$ and $t$ intersects,
and it is +1 (-1) if $\check{l}$,
after rotated 90 (-90) degrees anticlockwisely,
is equal to $t$.

We label by $P_N$ the canonical momentum of $N$: $[N,P_N]=i$.
The extended Hilbert space in the canonical momenta basis is
\begin{eqnarray}
|\{\pi_A(\ver)\in\mathbb{R}\}|\Delta\pi_A\equiv0\rangle\otimes|P_N\in\mathbb{R}\rangle,
\end{eqnarray}
where the Gauss law constraint realizes the gauge equivalence of $A$ in the $A$-basis.

\subsubsection{Unitary duality transformation}
\begin{figure}[h]
\centering
\includegraphics[width=8.8cm,pagebox=cropbox,clip]{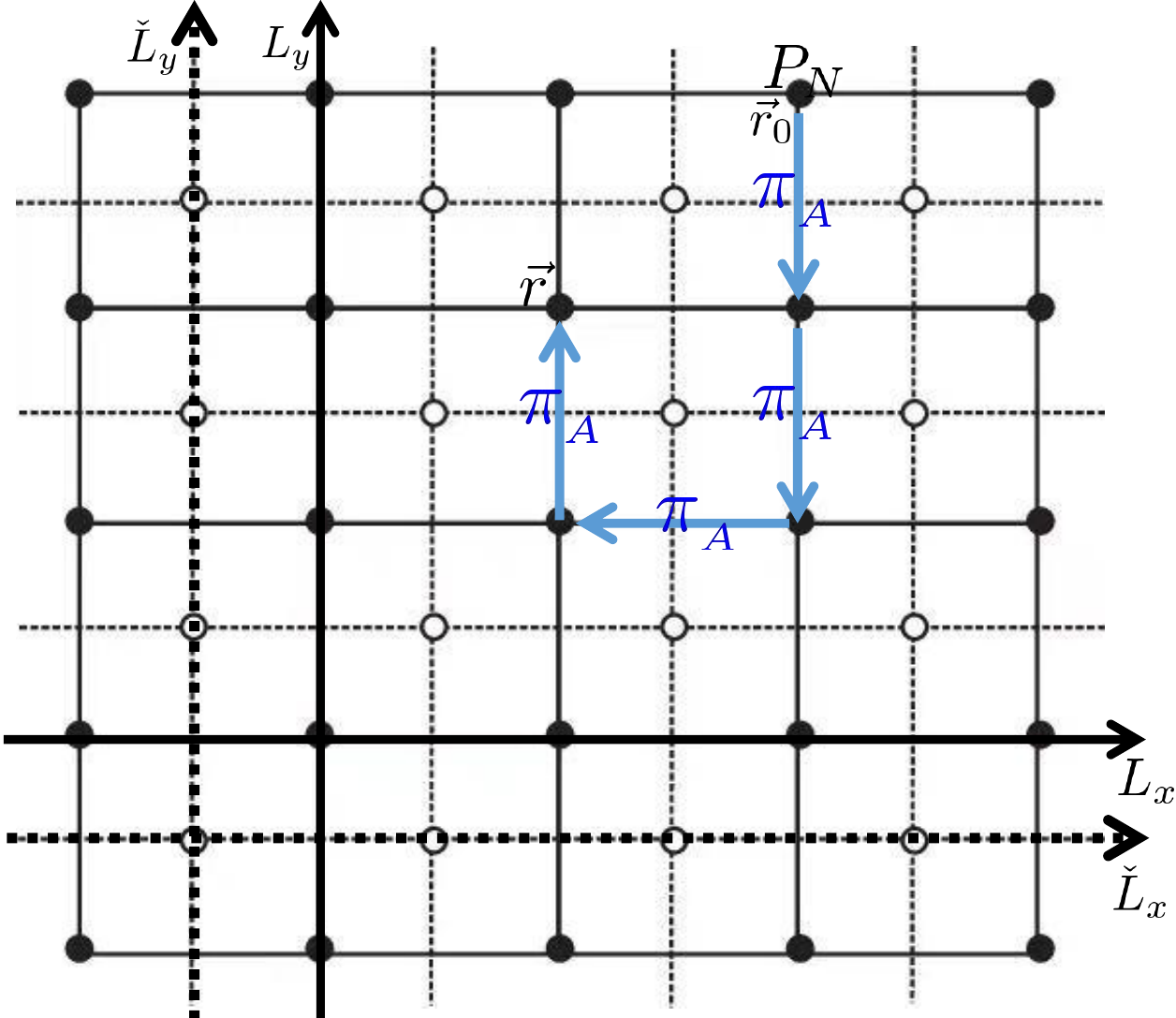}
\caption{Illustration of $AH\phi(\vec{r})AH^\dagger$.
For simplicity,
we take a path of $\pi_A$ from $\vec{r}_0$ to $\ver$ not crossing the boundary formed by $\check{L}_\mu$'s.}
\label{phi_dual}
\end{figure}
\begin{figure}[h]
\centering
\includegraphics[width=8.8cm,pagebox=cropbox,clip]{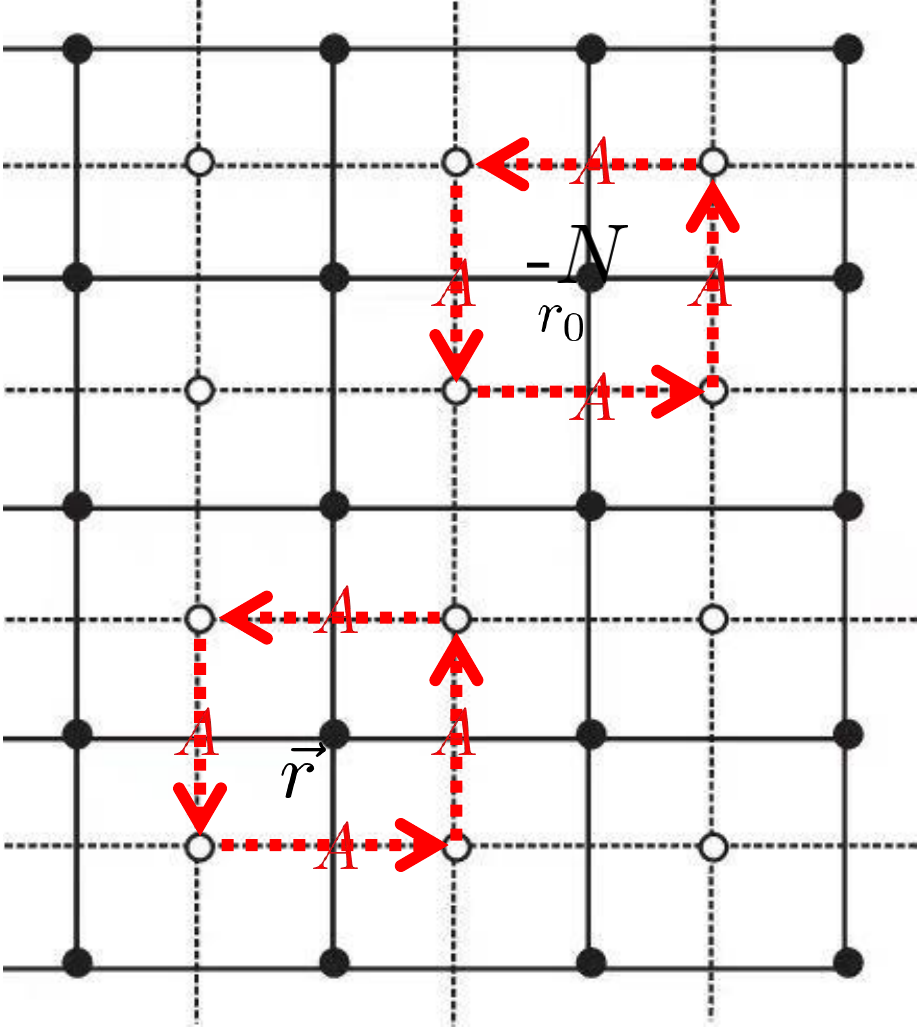}
\caption{Illustration of $AH\pi(\vec{r})AH^\dagger$.
When $\ver=\vec{r}_0$,
we need to include an additional $(-N)$.}
\label{pi_dual}
\end{figure}
We will label the duality transformation as $AH$ to denote ``Abelian-Higgs'',
although the current transformation is unitary in the extended Hilbert space:
\begin{eqnarray}\label{AH_2d}
\left\{\begin{array}{l}
AH\alpha_\mu AH^\dagger=-2\pi\sum_{\vec{t}\in L_\mu}\pi_A(\vec{t});\\
AHp_\alpha^\mu AH^\dagger=\frac{1}{2\pi}\sum_{\vec{t}\in\check{L}_\nu}A(\vec{t})\epsilon^{\nu\mu};\\
AH\phi(\ver)AH^\dagger=P_N+2\pi\sum_{\vec{t}\in l:\ver_0}^\ver\pi_A\left(\vec{t}\right);\\
AH\pi(\ver)AH^\dagger=\frac{1}{2\pi}[\Delta A](\ver)-N\delta_{\ver,\ver_0},\end{array}\right.
\end{eqnarray}
where $\ver_0$ is an arbitrarily chosen and fixed reference point on the original lattice.
%For convenience,
%$\ver_0$ is taken in the interior of the coordinate patch: $1<(\ver_0)_{x,y}<L_{x,y}$.
Here $l$ is an arbitrary path starting from $\ver_0$ to $\ver$,
not intersecting with $\check{L}_\mu$,
which is always possible~\footnote{Such a restriction,
together with the choice of $\check{L}_\mu$, can be avoided as in one dimension~(\ref{1_d_phi}),
but it will induce messy notations.}.
The arbitrariness requires that
\begin{eqnarray}
\Delta\pi_A\equiv0,
\end{eqnarray}
which is precisely the Gauss law imposed on the Hilbert space.
We can solve out the inverse:
\begin{eqnarray}\label{AH_2d_inv}
\left\{\begin{array}{l}
AH^\dagger NAH=-\sum_{\ver}\pi(\ver);\\
AH^\dagger P_NAH=\phi(\ver_0);\\
AH^\dagger \sum_{\vec{t}\in\check{L}_\mu}A(\vec{t})AH=2\pi\epsilon^{\mu\nu}p_\alpha^\nu;\\%+2\pi\sum_{\ver\in S:\partial S=\check{l}_\mu-\check{L}_\mu}\left[\pi(\ver)-\delta_{\ver,\ver_0}\sum_{\vec{s}}\pi(\vec{s})\right];\\
AH^\dagger [\Delta A](\ver)AH=2\pi\left[\pi(\ver)-\sum_{\vec{s}}\pi(\vec{s})\delta_{\ver,\ver_0}\right];\\
AH^\dagger \pi_AAH=\frac{1}{2\pi}\left(\Delta\phi-\alpha\right),
\end{array}\right.
\end{eqnarray}
where $\alpha$ is a link field such that $(\mu=x \text{ or }y)$
\begin{eqnarray}
\alpha(\ver,\ver+\mu)=\left[\delta_{(\ver)_x,L_x}\delta_{\mu,x}+(x\leftrightarrow y)\right]\alpha_\mu.
\end{eqnarray}
Consistently,
the Gauss law constraint is automatically satisfied since $\Delta\alpha=0$.
The above operator mappings naturally induce the corresponding Hilbert space transformations.
We have illustrated the duality transformation of $\phi(\ver)$ and $\pi(\ver)$ in FIG.~\ref{phi_dual} and FIG.~\ref{pi_dual}, respectively.

%\subsection{Case 2: $\mathbb{R}/2\pi-\mathbb{R}'$ duality}
%In this case,
%the scalar side has the Hilbert space as
%\begin{eqnarray}
%|\{\Delta\phi\in\mathbb{R}\},\exp[i\phi(\ver_0)]\in\text{U}(1)\rangle\otimes|\alpha_\mu\in\mathbb{R}\rangle,
%\end{eqnarray}
%in which the \textit{total} $\mathbb{R}$-charge is quantized in the canonical conjugate variable:
%\begin{eqnarray}
%|\{\pi_\ver\in\mathbb{R}\}_{\ver\neq\ver_0},\sum_\ver\pi_\ver\in\mathbb{Z}\rangle\otimes|p_\alpha^\mu\in\mathbb{R}\rangle.
%\end{eqnarray}
%
%The dual side is
%\begin{eqnarray}
%|\cdots\rangle\otimes\left[|\exp(iP_N)\in\text{U}(1)\rangle\text{ or }|N\in\mathbb{Z}\rangle\right],
%\end{eqnarray}
%which means the only effect on the dual side is the quantization of $N$.

\subsection{U(1)-$\mathbb{Z}$ duality}
Now,
we promote $\phi$ to be angle-valued:
\begin{eqnarray}
|\{\exp[i\phi(\ver)]\in\text{U}(1)\}\rangle\otimes|\exp(i\alpha_\mu)\in\text{U}(1)\rangle,
\end{eqnarray}
with canonical conjugate
as
\begin{eqnarray}
|\{\pi(\ver)\in\mathbb{Z}\}\rangle\otimes|p^\mu_\alpha\in\mathbb{Z}\rangle.
\end{eqnarray}
The dual side is the $\mathbb{Z}$-gauge height model:
\begin{eqnarray}
\begin{array}{l}|\{\sum_{\vec{t}\in\check{l}}A(\vec{t})\in2\pi\mathbb{Z}\}_{\check{l}}\rangle\otimes|N\in\mathbb{Z}\rangle,\end{array}
\end{eqnarray}
or its canonical conjugate:
\begin{eqnarray}
|\{\exp[i2\pi\pi_A(\ver)]\in\text{U}(1)\}|\exp(i2\pi\Delta\pi_A)\equiv1\rangle\otimes|\exp(iP_N)\in\text{U}(1)\rangle.
\end{eqnarray}
The unitary duality transformations are proper exponentiations of Eqs.~(\ref{AH_2d},\ref{AH_2d_inv}),
so we do not repeat them here as well as in the following discussion.

\subsection{$\mathbb{Z}$-U(1) duality}
In this case,
the dual part is the (extended) conventional lattice U(1)-gauge field with the Hilbert space as
\begin{eqnarray}
\begin{array}{l}|\{\prod_{\vec{t}\in\check{l}}\exp[iA(\vec{t})]\in\text{U}(1)\}_{\check{l}}\rangle\otimes|\exp(i2\pi N)\in\text{U}(1)\rangle,\end{array}
\end{eqnarray}
with canonical conjugate:
\begin{eqnarray}
|\{\pi_A\in\mathbb{Z}\}|\Delta\pi_A\equiv0\rangle\otimes|P_N\in\mathbb{Z}\rangle.
\end{eqnarray}
The other side is
the discrete height model:
\begin{eqnarray}
|\{\phi(\ver)\in2\pi\mathbb{Z}\}\rangle\otimes|\alpha_\mu\in2\pi\mathbb{Z}\rangle,
\end{eqnarray}
or the canonical conjugate
\begin{eqnarray}
\begin{array}{l}|\{\exp\left[i2\pi\pi(\ver)\right]\in \text{U}(1)\}\rangle\otimes|\exp(i2\pi p^\mu_\alpha)\in \text{U}(1)\rangle.\end{array}
\end{eqnarray}

\subsection{$\mathbb{Z}_n$-$\mathbb{Z}_n$ duality}
This case is completely paralleling to the one-dimensional case,
by substitutions of continuous variables by discrete matrix variables.
Therefore,
we leave it to the interested readers.

\section{Duality in arbitrary dimensions of arbitrary forms}\label{duality_arb}
The results in the preceding sections
can be naturally generalized to any dimension.
This section may be skipped in the first reading since it might use some technical notions unfamiliar to general audience.

First of all,
the lattice $X$ is a $d$-dimensional hypercubic lattice with PBC and $(p+1)$-cells or $(p+1)$-dimensional hyperplaquettes defined on it are denoted by $c_{p+1}$.
Their formal finite sum with $\mathbb{Z}$-coefficients form a module denoted by $C_{p+1}(X)$.
The dual lattice is $\check{X}$,
and $(p+1)$-cells thereon are labelled by $\check{c}_{p+1}$,
and the module is denoted as $C_{p+1}(\check{X})$.
The homology group
\begin{eqnarray}
&&H_{p+1}(X)\cong H_{d-p-1}(\check{X})\cong\left(\mathbb{Z}\right)^{\cc{d}{p+1}},
\end{eqnarray}
and let us choose by hand a set of \textit{representatives} of the generators of $H_{p+1}(X)$ and $H_{d-p-1}(\check{X})$ as
\begin{eqnarray}
&&\gamma_{i}\in C_{p+1}(X),\,\,\check{\gamma}^i\in C_{d-p-1}(\check{X}),\,\,i=1,\cdots,\cc{d}{p+1};\\
&\text{s.t., }&\text{Int}(\gamma_i,\check{\gamma}^j)=\delta^j_i.
\end{eqnarray}
Similarly,
we also choose a set of representatives of generators of $H_{p}(X)$ and $H_{d-p}(\check{X})$:
\begin{eqnarray}
&&\eta_{i}\in C_{p}(X),\,\,\check{\eta}^i\in C_{d-p}(\check{X}),\,\,i=1,\cdots,\cc{d}{p};\\
&\text{s.t., }&\text{Int}(\check{\eta}^j,\eta_i)=\delta^j_i.
\end{eqnarray}

{\bf Remark: }We have chosen a set of chains to represent the homological classes,
so the following discussions will potentially \textit{unnatural} in the sense that they might depend on this choice and we will remind you of these issues.

On one side of the duality,
we have the following degrees of freedom:
\begin{eqnarray}
|B\in C^{p}(X,G)/B^p(X,G),\,\,M\in H^{p+1}(X,G)\rangle,
\end{eqnarray}
or their canonical momenta formally staying on the dual lattice $\check{X}$
\begin{eqnarray}
|\pi_B\in Z^{d-p}(\check{X},\check{G}),\,\,P_M\in H^{d-p-1}(\check{X},\check{G})\rangle.
\end{eqnarray}
Here $Z^{\cdot}$ denotes the cocycle group,
$B^{\cdot}$ the coboundary group,
and $H^{\cdot}$ the cohomology group.
To our current interest,
we consider
\begin{eqnarray}
(G,\check{G})=(\mathbb{R},\mathbb{R}),\,\,(\text{U}(1),\mathbb{Z}),\,\,(\mathbb{Z},\text{U}(1)),\,\, \text{or } (\mathbb{Z}_n,\mathbb{Z}_n).
\end{eqnarray}

The dual theory space takes form as
\begin{eqnarray}
|A\in C^{d-p-1}(\check{X},\check{G})/B^{d-p-1}(\check{X},\check{G}),\,\,N\in H^{d-p}(\check{X},\check{G})\rangle,
\end{eqnarray}
or their canonical momenta formally staying on the original lattice $X$
\begin{eqnarray}
|\pi_A\in Z^{p+1}(X,G),\,\,P_N\in H^p(X,G)\rangle.
\end{eqnarray}

Due to the universal coefficient theorem,
$N$ ($P_N$) is completely defined once its integral on all the $\gamma_i$'s ($\check{\gamma}^i$'s) is known.
Their defining canonical structure are
\begin{eqnarray}
&&\left[\oint_{\gamma_i}M,\oint_{\check{\gamma}^j}P_M\right]=i\text{Int}(\gamma_i,\check{\gamma}^j)=\sqrt{-1}\delta^j_i,\\
&&\left[\oint_{\check{\eta}^j}N,\oint_{{\eta}_i}P_N\right]=i\text{Int}(\check{\eta}^j,\eta_i)=\sqrt{-1}\delta^j_i.
\end{eqnarray}
The other dynamical fields,
on arbitrary plaquettes, satisfy,
\begin{eqnarray}
&&\left[B(c_p),\pi_B(\check{c}_{d-p})\right]=i\text{Int}(c_p,\check{c}_{d-p}),\\
&&\left[A(\check{c}_{d-p-1}),\pi_A(c_{p+1})\right]=i\text{Int}(\check{c}_{d-p-1},c_{p+1}).
\end{eqnarray}

The duality transformation ``$Dual$'' takes the form as:
\begin{eqnarray}\label{dual_1}
\left\{\begin{array}{l}Dual^\dagger \oint_{\check{\eta}^i}N Dual=-2\pi\oint_{\check{\eta}^i}\pi_B;\\
Dual^\dagger \oint_{\eta_i} P_N Dual=\frac{1}{2\pi}\oint_{\eta_i}B;\\
Dual^\dagger \oint_{\check{\gamma}^i}A \,\,Dual=2\pi \oint_{\check{\gamma}^i}P_M;\\
Dual^\dagger [\Delta A-2\pi N] Dual=2\pi\pi_B;\\
Dual^\dagger\pi_A Dual=\frac{1}{2\pi}[\Delta B-2\pi M].\end{array}\right.
\end{eqnarray}
The inverse of the transformation is
\begin{eqnarray}\label{dual_2}
\left\{\begin{array}{l}Dual \oint_{\gamma_i}M Dual^\dagger=-2\pi\oint_{\gamma_i}\pi_A;\\
Dual \oint_{\check{\gamma}^i} P_M Dual^\dagger=\frac{1}{2\pi}\oint_{\check{\gamma}^i}A;\\
Dual \oint_{\eta_i}B \,\,Dual^\dagger=2\pi \oint_{\eta_i}P_N;\\
Dual [\Delta B-2\pi M] Dual^\dagger=2\pi\pi_A;\\
Dual\,\pi_B Dual^\dagger=\frac{1}{2\pi}[\Delta A-2\pi N].\end{array}\right.
\end{eqnarray}

It should be noted that
the dualization of $N$ or $M$ is independent on the choice of $\gamma_i$'s or $\check{\gamma}^i$ because $\Delta\pi_A=0$ and $\Delta \pi_B=0$.
In contrast,
the dualization of $P_N$ or $P_M$ relies on the choice of $\check{\gamma}^i$, $\gamma_i$, $\check{\eta}^i$ and $\eta_i$.
In the previous sections,
such dependences are through the choices of $L_\mu$, $L_{\check{\mu}}$,
and $\ver_0$.

\section{Exact UV realizations of field-theoretical dualities}\label{field_dual}
In this section,
we will see how the field-theoretical dualities in the IR can be realized on the lattice in the UV.

Let us first briefly review the IR physics with $d=1$ and $p=0$.
In such a field theory,
the field configuration of $\phi(x)$ is continuous,
so we could not have singularity like vortices.
However,
$\phi(x)$ should be treated as an angle-valued quantity.
How to solve such a contradiction;
we know that the vortices are inevitable in the spectrum once we have an angle-valued field.

The solution is that the lattice realization need to be a \textit{non-local} theory.
We observe that
the field theory locally behaves the same as the $\mathbb{R}$-valued $\phi(x)$ and the $\mathbb{R}/2\pi\mathbb{Z}$ property of $\phi(x)$ is reflected in the boundary condition.
We recall that the boundary condition,
when the Hilbert space is extended,
is dynamically determined by $\alpha$.
Thus,
the pre-dualized side of the duality has the Hilbert space as:
\begin{eqnarray}
\frac{|\{\phi_j\in\mathbb{R}\}_{j=1,\cdots,L}\rangle}{\{\phi_j\}\sim\{\phi_j\}+2\pi}\otimes|\alpha\in 2\pi\mathbb{Z}\rangle,
\end{eqnarray}
which is a non-local theory due to the ``gauge'' redundancy by a \textit{global} symmetry transformation.
Such a gauge redundancy can be imposed by that
the global transformation by $2\pi$ is ``doing nothing'',
or
\begin{eqnarray}
\exp\left(i2\pi\sum_j\pi_j\right)=1,
\end{eqnarray}
or equivalently,
$\sum_j\pi_j\in\mathbb{Z}$.
Therefore,
the Hilbert space in the canonical momentum space takes the form as
\begin{eqnarray}
\left|\{\pi_j\in\mathbb{R}\}_{j=1,\cdots,L}\left|\begin{array}{l}\sum_{j}\end{array}\pi_j\in2\pi\mathbb{Z}\right.\right\rangle\otimes|p_\alpha\in\mathbb{R}/\mathbb{Z}\rangle,
\end{eqnarray}
where the ``gauge'' redundancy of $p_\alpha$ is due to the quantization of $\alpha/2\pi$.

The dual side has the same structure:
\begin{eqnarray}
\frac{|\{h_{j-1/2}\in\mathbb{R}\}_{j=1,\cdots,L}\rangle}{\{h_{j-1/2}\}\sim\{h_{j-1/2}\}+2\pi}\otimes|\beta\in2\pi\mathbb{Z}\rangle,
\end{eqnarray}
or
\begin{eqnarray}
\left|\{\pi_{h,j-1/2}\in\mathbb{R}\}_{j=1,\cdots,L}\left|\begin{array}{l}\sum_{j}\end{array}\pi_{h,j-1/2}\in2\pi\mathbb{Z}\right.\right\rangle\otimes|p_\beta\in\mathbb{R}/\mathbb{Z}\rangle.
\end{eqnarray}
The duality transformations are formally the same as before in Eqs.~(\ref{1_d_phi},\ref{1_d_h}) except for that those relations associated to $p_\alpha$ and $p_\beta$ should be properly exponentiated as $\exp(2\pi ip_\alpha)$ and $\exp(2\pi ip_\beta)$.

In general cases of $(d,p)$,
the pre-dualized side has the Hilbert space as
\begin{eqnarray}
|B\in {C}^{p}(X,\mathbb{R})/Z^p(X,2\pi\mathbb{\mathbb{Z}}),\,\,M\in H^{p+1}(X,2\pi\mathbb{Z})\rangle,
\end{eqnarray}
where one should note the quotient of $C^{\cdot}$ is by $Z^{\cdot}$ now rather than $B^{\cdot}$ before
and the coefficient group therein is also changed.

The canonical momentum correspondence is:
\begin{eqnarray}
|\pi_B\in Z^{d-p}(\check{X},\mathbb{R}),\,\,P_M\in H^{d-p-1}(\check{X},\mathbb{R}/\mathbb{Z})|\begin{array}{l}\oint_{\check{\eta}^i}\pi_B\in\mathbb{Z}\end{array}\rangle.
\end{eqnarray}
As a technical digression,
the condition $\oint\pi_B\in\mathbb{Z}$ can be alternatively rephrased as that $\pi_B$ can represent an element of $H^{d-p}(\check{X},\mathbb{Z})$  
after some modification by terms in $B^{d-p}(\check{X},\mathbb{R})$.

The dual side has the same structure:

\begin{eqnarray}
|A\in C^{d-p-1}(\check{X},\mathbb{R})/Z^{d-p-1}(\check{X},2\pi\mathbb{Z}),\,\,N\in H^{d-p}(\check{X},2\pi\mathbb{Z})\rangle,
\end{eqnarray}
or its canonical conjugate:
\begin{eqnarray}
|\pi_A\in Z^{p+1}(X,\mathbb{R}),\,\,P_N\in H^p(X,\mathbb{R}/\mathbb{Z})|\begin{array}{l}\oint_{\gamma_i}\end{array}\pi_A\in\mathbb{Z}\rangle.
\end{eqnarray}
The dual transformations are still Eqs.~(\ref{dual_1},\ref{dual_2}),
in which those equations related to $P_N$ and $P_M$ must be exponentiated properly.

One final remark is that the so-called self-dualities occurs at
$p=d-p-1$ or $2p=d-1$.
Especially,
$(d,p)=(1,0)$ corresponds to ``$\varphi$-$\sigma$'' duality,
and
$(d,p)=(3,1)$ the EM duality.
The above dualities can be also understood in the Euclidean space-time lattice by a modified Villain form~\cite{Gorantla:2021aa}.

\section{Duality of LSM arguments in two dimensions}
Unlike in one dimension,
we cannot prove LSM theorem for rotor models or gauge fields in two dimensions or higher rigorously,
although there are some reasonable arguments~\cite{Lieb:1961aa,Affleck:1988,Oshikawa:2000aa,Kobayashi:2018aa}.
Therefore,
we will call them ``{\bf Statements}'' and try to deduce their dualities.

{\bf Statement 1: }\textit{LSM theorem for quantum rotor models in two dimensions --- }
If a quantum rotor Hamiltonian in two dimensions respects lattice translation symmetry $T_{x,y}$:
\begin{eqnarray}
T_\mu\exp[i\phi(\ver)]T_\mu^{-1}=\exp[i\phi(\ver+\mu)];\,\,T_\mu\pi(\ver)T_\mu^{-1}=\pi(\ver+\mu),\,\,\mu=x,y,
\end{eqnarray}
under PBC,
and U$(1)$-rotational symmetry generated by
\begin{eqnarray}
\prod_\ver\exp[i\theta\pi(\ver)],\,\,\theta\in[0,2\pi),
\end{eqnarray}
then,
as $L_{x,y}\rightarrow\infty$, there must exist multiple lowest-lying energy eigenstates \textit{within} a fixed U$(1)$-charge sector with $p/q$-fractional charge per unit cell
\begin{eqnarray}
Q=\sum_{\ver}\pi(\ver)=\frac{p}{q}L_xL_y.
\end{eqnarray}
{\bf Remarks: }
We have suppressed the ``well-defined thermodynamic limit'' due to two reasons:

(1) This condition should be satisfied by generic physical short-range interacting systems which are extensible;

(2) {\bf Statement 1} cannot be proven as rigorously as its one-dimensional reduction,
but it can be argued through a flux-insertion argument which adopts a no-gap-closing assumption.
It is unclear know how ``well-defined thermodynamic limit'' is essential to this statement.

Then we can propose its dual:

{\bf Statement $\check{1}$ --- }\textit{LSM theorem of quantum $\mathbb{Z}$-gauge models in two dimensions:}
If a quantum $\mathbb{Z}$-gauge Hamiltonian respects ``modulating'' lattice translation symmetry $T_{x,y}^{(p/q)}$ with any $q_{x,y}$ such that $q_xq_y=q$ and $q_\mu$ is divisible by $L_\mu$:
\begin{eqnarray}\label{mod_transl_2}
\left\{\begin{array}{l}T^{(p/q)}_x A(\check{\ver},\check{\ver}+x){T^{(p/q)}_x}^{-1}=A(\check{\ver}+x,\check{\ver}+x+x);\\
T^{(p/q)}_y A(\check{\ver},\check{\ver}+x){T^{(p/q)}_y}^{-1}=A(\check{\ver}+y,\check{\ver}+x+y)-p\delta_{\check{\ver}_x=1\text{ mod }q_x}\delta_{\check{\ver}_y=1\text{ mod }q_y};\\
T^{(p/q)}_x A(\check{\ver},\check{\ver}+y){T^{(p/q)}_x}^{-1}=A(\check{\ver}+x,\check{\ver}+y+x)+p\delta_{\check{\ver}_x=1\text{ mod }q_x}\delta_{\check{\ver}_y=1\text{ mod }q_y};\\
T^{(p/q)}_y A(\check{\ver},\check{\ver}+y){T^{(p/q)}_y}^{-1}=A(\check{\ver}+y,\check{\ver}+y+y),\end{array}\right.
\end{eqnarray}
with $\pi_A$'s invariant ($\check{r}$'s are dual lattice sites).
It should be noted that the above transformations are formal and they are well-defined only after inserted into the gauge-invariant variables,
e.g., exponentiated Wilson lines.
The Hamiltonian is also required to preserve $\mathbb{Z}$-raising 1-form symmetry generated by
\begin{eqnarray}
\prod_{t\in l}^L\exp[im\pi_A(t)],\,\,m\in\mathbb{Z},
\end{eqnarray}
where $l$ is any \textit{closed} loop.
As $L_{x,y}\rightarrow\infty$, there must exist multiple lowest-lying energy eigenstates \textit{within any} $\mathbb{Z}$-symmetry charge Hilbert subspace.

\textit{Sketch of the ``proof'' (since Statement 1 has not been rigorously proved):}
The dualization is almost similar to the one-dimensional correspondence;
we first ``accumulate'' by a unitary transformation $U_{(p/q)}$ all the twistings in Eq.~(\ref{mod_transl_2}) to the single plaquette centered at $\ver_0$:
\begin{eqnarray}
T_\mu\equiv T^{(p/q)}_\mu U^\dagger_{(p/q)},
\end{eqnarray}
satisfying
\begin{eqnarray}
&&T_\mu A(\check{\ver},\check{\ver}+\nu)T_\mu^{-1}=A(\check{\ver}+\mu,\check{\ver}+\nu+\mu)+2\pi \left(-\frac{p}{q}L_xL_y\right)\epsilon_{\nu\mu}\delta_{\check{\ver},\ver_0-(\mu+\nu)/2},\,\,(\mu,\nu=x,y);\nonumber\\
&&T_\mu\exp[i\pi_A(\ver)]T_\mu^{-1}=\exp[i\pi_A(\ver+\mu)].
\end{eqnarray}

Then in the extended Hilbert space: (For simplicity, we will assume $\ver_0$ is distant from $\check{L}_\mu$'s at least by two lattice constants.)
\begin{eqnarray}
\left\{\begin{array}{l}\mT_\mu N \mT_\mu^{-1}=N;\\
\mT_\mu\exp(iP_N)\mT_\mu^{-1}=\exp(iP_N)\exp[i\pi_A(\ver_0,\ver_0+\mu)];\\
\mT_\mu A(\check{\ver},\check{\ver}+\nu)\mT_\mu^{-1}=A(\check{\ver}+\mu,\check{\ver}+\nu+\mu)+2\pi N\epsilon_{\nu\mu}\delta_{\check{\ver},\ver_0-(\mu+\nu)/2};\,\,(\mu,\nu=x,y)\\
\mT_\mu\exp[i\pi_A(\ver)]\mT_\mu^{-1}=\exp[i\pi_A(\ver+\mu)],\end{array}\right.
\end{eqnarray}
and
\begin{eqnarray}
\left\{\begin{array}{l}(\mT_\mu\circ AH)\exp(i\hat{\alpha}_\nu)(\mT_\mu\circ AH)^{-1}=\exp(i\hat{\alpha}_\nu);\\
(\mT_\mu\circ AH)p^\nu_\alpha(\mT_\mu\circ AH)^{-1}=p^\nu_\alpha+2\pi\sum_{\ver\in L_\rho}\pi(\ver)\epsilon_{\rho\nu}\delta_\mu^\nu;\\
(\mT_\mu\circ AH)\exp[i\phi(\ver)](\mT_\mu\circ AH)^{-1}=\exp[i\phi(\ver)]\exp(i\hat{\alpha}_\mu\delta_{\ver_\mu,L_\mu});\\
(\mT_\mu\circ AH)\pi(\ver)(\mT_\mu\circ AH)^{-1}=\pi(\ver+\mu),
\end{array}\right.
\end{eqnarray}
where $L_\rho$'s are defined in FIG.~\ref{phi_dual} and we have used the fact or gauge choice that $\ver_0$ is distant from $\check{L}_\mu$'s by at least two lattice constants.
As before,
the detailed form of the translation rules of $\exp(iP_N)$ or $p^\nu_\alpha$ is not essential,
but their good behavior on both original and dual side is important so that the {\bf Statement 1} can be used.

Comparing $\mT_\mu$ and $T_\mu$,
we find that
\begin{eqnarray}
N=-\frac{p}{q}L_xL_y,
\end{eqnarray} 
which gives
\begin{eqnarray}
Q=AH^\dagger (-N) AH=\frac{p}{q}L_xL_y,
\end{eqnarray}
so {\bf Statement $\check{1}$} follows {\bf Statement $1$}.

Conversely starting from the gauge field,
we have the following statement~\cite{Kobayashi:2018aa}:

{\bf Statement 2: }\textit{LSM theorem for U(1)-gauge theory in two dimensions --- }
If a two-dimensional quantum U(1)-gauge Hamiltonian  respects 1-form U(1)-symmetry generated by
\begin{eqnarray}
\exp\left[i\sum_{t\in{L}_\mu}i\theta \pi_A(t)\right],\,\,(\mu=x,y;\theta\in[0,2\pi)),
\end{eqnarray}
and one of the lattice translation symmetry,
e.g., $T_x$,
then it possesses multiple low-lying energy eigenstates within 1-form U(1) fractional-charge Hilbert subspace along $x$ axis:
\begin{eqnarray}
\sum_{t\in{L}_x} \pi_A(t)=\frac{p}{q}L_x.
\end{eqnarray}

{\bf Remark: }{\bf Statement 2} can be justified by a flux-insertion argument~\cite{Kobayashi:2018aa},
or alternatively,
when we compactify $y$-axis, it would reduce to {\bf Theorem 1},
although this dimensional reduction cannot work as a rigorous proof due to a thermodynamic-limit order problem~\cite{Lieb:1961aa,Affleck:1988}.
Other arguments without thermodynamic issue can be found in~\cite{Yao:2020PRX,Yao:2021aa}.

Then we can dualize the above Statement to obtain:

{\bf Statement $\check{2}$: }\textit{LSM theorem for quantum $\mathbb{Z}$-height model in two dimensions --- }
If a quantum $\mathbb{Z}$-height Hamiltonian respects ``modulating'' lattice translation symmetry $T_{p/q}$ only along $x$-axis:
\begin{eqnarray}
T_{p/q}\phi_{\ver}T_{p/q}^{-1}=\phi_{\ver+x}+p\delta_{\ver_x=1\text{ mod }q};\,\,T_{p/q}\pi_{\ver}T_{p/q}^{-1}=\pi_{\ver+x},
\end{eqnarray}
under PBC,
and an onsite $\mathbb{Z}$-raising symmetry generated by
\begin{eqnarray}
\prod_{\ver}\exp(im\pi_{\ver}),\,\,m\in\mathbb{Z},
\end{eqnarray}
then there must exist multiple lowest-lying energy eigenstates \textit{within any} $\mathbb{Z}$-symmetry charge Hilbert subspace.

\textit{Sketch of the ``proof'': }The proof is similar to the one-dimensional analog.
Obviously,
the dualization,
after a dimensional reduction,
becomes that between {\bf Theorems $1$ and $\check{1}$}.
A systematic approach is to accumulate the twistings, to extend the Hilbert space,
and to use the extended translation symmetry as designed above.

\section{Dualization of ingappabilities in arbitrary dimensions}

We can formally generalize the results so far in $d=1$ and $d=2$ toward arbitrary dimensions and forms.

{\bf Statement ${(p+1)=1,\cdots,d}$: }\textit{LSM theorem for $p$-form U(1)-gauge theory --- }
In $d$-dimensional lattice,
there exists $d$ types of LSM-type statements;
for a $p$-form U(1)-gauge theory $(p=0,\cdots,d-1)$,
there exists multiple low-lying energy eigenstates if it possesses $p$-form U(1)-symmetry and translation symmetry along $(d-p)$ of axes within a fractional $p$-form charge density Hilbert subspace along this $(d-p)$-dimensional hypersurface.

Their dualizations turns out to be:

{\bf Statement ${\check{(p+1)}=1,\cdots,d}$: }\textit{LSM theorem for $(d-p-1)$-form $\mathbb{Z}$-gauge theory --- }
If the $\mathbb{Z}$-gauge theory possesses modulating translation symmetry generalizing Eqs.~(\ref{mod_transl_1},\ref{mod_transl_2}) and respects $(d-p-1)$-form $\mathbb{Z}$-symmetry,
there exists multiple low-lying energy eigenstates within any $(d-p-1)$-form $\mathbb{Z}$-charge eigenspace.

Especially,
EM duality induced LSM theorem is the case of $d=3$ and $p=1$.

\section{Discussions on the non-invertibility}\label{discussion}
In this work,
we systematically dualize the LSM-type theorems through KW duality and its higher-dimensional and higher-form generalizations.
We conclude this paper by considering the duality transformation between the physical Hilbert spaces.
We first take $d=1$ as illustration;
let us restrict the physical Hilbert space on pre-dualized side within the subspace with $\hat{\alpha}=\alpha_0$ and the dual side with fixed $\hat{\beta}=\beta_0$,
then
the conventional (non-unitary) duality transformation is
\begin{eqnarray}
kw&\equiv &\text{Proj}_{\beta_0}\circ KW \circ\text{Proj}_{\alpha_0},
\end{eqnarray}
where Proj's are projection operators.
Thence
\begin{eqnarray}
&&kw^\dagger \circ kw\propto\text{``}\int d\theta\,\,\text{''}\exp(-i\sum_k\theta\pi_k)\exp(-i\theta\beta_0),\\
&&kw \circ kw^\dagger\propto\text{``}\begin{array}{l}\sum_{\check{\theta}}\end{array}\text{''}\exp(-i\sum_k\check{\theta}\pi_{h,k+1/2})\exp(-i\check{\theta}\alpha_0),
\end{eqnarray}
up to a proper normalization constant and the formal notations of integration or summation ``$\int d\theta$'' and ``$\sum_{\check{\theta}}$'' are determined by the Hilbert spaces involved,
e.g., $\theta$ is integrated over $[0,2\pi)$ and $\check{\theta}$ is summed over integers in the case of angle-valued $\phi\sim\phi+2\pi$ on the pre-dual side.

For AH duality,
the physical space is a subspace with a fixed $\hat{\alpha}_{x,y}=\alpha_{0x,y}$ value on one side,
and a fixed $\hat{N}=N_0$ on the dual side.
Similarly,
\begin{eqnarray}
ah\equiv\text{Proj}_{N_0}\circ AH\circ\text{Proj}_{\alpha_{0\mu}},
\end{eqnarray}
which gives that
\begin{eqnarray}
&&ah^\dagger \circ ah=\text{``}\int d\theta\text{\,\,''}\exp(-i\sum_k\theta\pi_k)\exp(-i\theta N_0),\nonumber\\
&&ah\circ ah^\dagger=\prod_{\mu=x,y}\text{``}\begin{array}{l}\sum_{\check{\theta}_\mu}\end{array}\text{''}\exp\left[-i\sum_{t\in L_\mu}\check{\theta}_\mu\pi_A(t)\right]\exp(-i\check{\theta}_\mu\alpha_\mu).
\end{eqnarray}

The calculation is directly generalizable to higher-dimensional and higher-form cases for $Dual$ or $dual$. 

\section{Acknowledgements}
The author thank Linhao Li, Masaki Oshikawa, Zijian Xiong, and Yunqin Zheng for useful discussions,
and the sponsorship from Yangyang Development Fund and Xiaomi Young Scholars Program.

%\bibliography{bib}

%merlin.mbs apsrev4-1.bst 2010-07-25 4.21a (PWD, AO, DPC) hacked
%Control: key (0)
%Control: author (8) initials jnrlst
%Control: editor formatted (1) identically to author
%Control: production of article title (-1) disabled
%Control: page (0) single
%Control: year (1) truncated
%Control: production of eprint (0) enabled
%

\end{document}